# Autonomous Quantum Transport Measurements of 2D Semiconductors by an AI Agent

Brandon Bauer[1], Matthew Whalen[1], Kenji Watanabe[2], Takashi Taniguchi[2], John Q. Xiao[1], Tao E. Li[1*], Wenjin Zhao[1*]

[1]Department of Physics and Astronomy, University of Delaware, Newark, DE, USA
[2]National Institute for Materials Science, 1-1 Namiki, 305-0044 Tsukuba, Japan

Email: wzhao@udel.edu; taoeli@udel.edu

**Abstract:**

Artificial-intelligence (AI) agents are beginning to enter experimental laboratories, automating experiments and accelerating scientific discovery. Herein, we introduce an AI-driven workflow in which an AI agent performs multi-step, multi-day quantum transport measurements end-to-end. Specifically, given brief instructions, the agent starts by planning the multi-step measurements, then safely operates the cryogenic instruments, analyzes the data, and concludes with a final report. We demonstrate this AI workflow on multiple monolayer and bilayer $MoS_2$ devices. Through autonomous measurement campaigns lasting up to six days, the agent determined the conduction-band spin–orbit coupling energy in monolayer $MoS_2$, and mapped a layer- and valley-resolved phase diagram in bilayer $MoS_2$. This experimental workflow is implemented through the FermiLink agent harness, which emphasizes instrumental safety and the reliability of the measurement and analysis. The framework is general and can be readily adapted to other types of experiments, representing a step toward self-driving laboratories.

## Introduction

Quantum transport measurements are central to the discovery of electronic phases in quantum materials, including superconductors, topological insulators, and strongly correlated insulators [1–4]. Understanding these materials often requires high-dimensional datasets, which typically depend on experimental parameters ranging from multiple gate voltages and magnetic fields to temperature. Acquiring and interpreting such datasets is time-consuming and requires substantial expertise in condensed-matter physics. The advance of large-language models (LLMs) and artificial intelligence (AI) agents raises the possibility of automating tasks normally performed by experimentalists [5–16]. However, important barriers remain. Specifically, as quantum material devices are delicate and cryogenic instruments are fragile and valuable, unreliable agent actions can compromise both the experiment and equipment. While literature for reporting single-step or human-in-the-loop experimental measurements is emerging, end-to-end AI-enabled multi-step quantum transport measurements in the absence of human intervention are still missing [15,16].

Here, we introduce a design principle for robust, agent-enabled, multistep, end-to-end quantum transport measurements. We start by examining the capacity of our agent workflow on *simulated* measurements of two-dimensional (2D) materials. With satisfactory performance, we utilized this agent framework by measuring two different types of $MoS_2$ devices from end-to-end with cryogenic instruments. For monolayer $MoS_2$ devices, the agent successfully determined the conduction-band spin–orbit coupling (SOC) energy, whereas for bilayer $MoS_2$ devices, it mapped the layer- and valley-resolved phase diagram. As this agent workflow directly generates high-quality quantum transport measurement reports from very brief initial instruction, we believe this AI automation attempt may provide a valuable resource for the quantum materials community.

**Autonomous Experimentation Framework**

Published papers and commercially available LLMs usually do not contain detailed information on how to perform a quantum transport study and the associated decision tree regarding when to pursue or abandon an experimental step. Moreover, because one quantum transport measurement may take hours or days to finish, this duration may exceed a typical session of an agent run using Codex or Claude Code, leading to unstable outcomes during multi-step measurements using commercially available AI agents. Aiming for autonomous agent-enabled multi-step quantum transport measurements, we first established an experimental loop workflow within the recently developed FermiLink agent harness [17]. This agent harness robustly enables multi-step experimental measurements, where the harness fully controls polling of instrumental status during long-waiting measurements, while the underlying AI agents, such as Codex and Claude Code, determine the next measurement commands and post-processing for each experimental step.

Then, we create an initial template for hypothesis-driven Agent Skills based on our experience, which provides detailed domain knowledge and hard-coded controlling and post-processing scripts for previous transport measurements on 2D quantum materials. Since the initial template is not robust enough to enable multi-step agent measurements, we use a small set of existing measurement data and apply the FermiLink harness to perform *simulated* measurements (which we call dry runs) and update the Agent Skills iteratively (Fig. 1b), until the Agent Skills become sufficiently stable for enabling end-to-end multi-step autonomous measurements (Supplementary Figure 1).

Fig. 1a provides a detailed multi-step workflow for autonomous quantum transport measurements, given the established FermiLink harness and converged Agent Skills. After a 2D materials device is loaded in a dilution refrigerator and the user provides a brief measurement instruction (e.g., "*Measure the conduction-band spin-orbit coupling energy of monolayer $MoS_2$*"), the FermiLink harness (*i*) assigns a measurement agent to design a detailed parameter sweep for a focused task on the dilution refrigerator and electronics. After the measurement agent submits a single-step long-duration measurement, FermiLink (*ii*) polls the experimental status in the absence of AI agents and (*iii*) then handles the finished experimental data and project memory to a new analysis agent for reasoning and post-processing. Each complete step from (*i*) to (*iii*) answers a focused scientific question, and the iterative loop of procedures ensures that the multi-step, multi-day quantum transport measurement, which ends with a comprehensive report, can be accomplished autonomously without human intervention. In this work, Codex with the LLM model gpt-5.5-xhigh is used as the underlying measurement and analysis agent, although the FermiLink harness can also connect to other AI agents such as Claude Code and OpenCode.

One key feature of our agent framework is protecting both the 2D materials devices and the instrument. For example, all hardware access is mediated by a frozen acquisition Python script rather than controlled directly by AI agents. Hardware safety constraints such as gate voltage limits, sweep rates, magnetic field, and temperature range are hard-coded in the Python script, prohibiting the agent from modifying them. This hard safety gate increases our confidence in using AI agents for end-to-end experiments.

## $MoS_2$ Devices and Electronic Structure

We use $MoS_2$ devices to demonstrate the agent-enabled autonomous quantum transport measurements. $MoS_2$ belongs to group-VI transition metal dichalcogenides (TMDs), $MX_2$ (M = Mo or W; X = S, Se, or Te) [18–20]. As shown in Fig. 2a, TMDs have a hexagonal crystal structure consisting of a transition metal layer sandwiched between two chalcogen layers in trigonal prismatic coordination geometry. These materials are semiconductors with band gaps typically in the range of 1–2 eV. Monolayer TMDs are direct-gap semiconductors, with band edges located at the $K$ and $K'$ valleys [21] (Fig. 2b). Near these valley extrema, the conduction and valence band edges exhibit parabolic dispersions. Because inversion symmetry is broken in monolayer TMDs, SOC lifts the spin degeneracy. The spin polarization is out-of-plane, with spin splitting typically on the order of hundreds of meV in the valence band and several to tens of meV in the conduction band [22]. We focus on the conduction band in the rest of the study.

In bilayer $MoS_2$, the top layer stacks antiparallel to the bottom layer, and the inversion symmetry is recovered. At $K$ and $K'$ valleys, the spin-up and spin-down bands become degenerate. In addition, the six $Q$ valleys, located roughly midway between Γ and K, can approach or even drop below the $K$ and $K'$ valleys in the conduction band [23–25]. Moreover, the interlayer tunneling at the $K$ and $K'$ valleys is small due to the momentum mismatch of the band with the same spin, so the layer degree of freedom enters. At a small perpendicular electric field, the $K$ and $K'$ valleys of both layers are nearly degenerate, whereas a finite electric field polarizes carriers onto one layer [26,27]. Distinguishing these scenarios from quantum transport requires combining quantum oscillation frequencies, Hall densities, and their evolution with gate voltages [28–32].

The $MoS_2$ device has a Hall bar shape with four gates as shown in Figs. 2c and 2d. The top and bottom gates control the carrier density and electric field independently. The combination of contact and split gates enables transport measurements at low perpendicular electric field and carrier density. To form low-resistance ohmic contacts to the conduction band of $MoS_2$, we use silver for the electrode because it has a low work function of about 4.2 eV and matches well with the conduction band [33]. Fig. 2e shows a representative *I-V* characteristic at the base temperature (10 mK) of the dilution refrigerator. A clear metal-to-insulator transition is also observed (Supplementary Figure 2). Unless otherwise specified, all measurements reported below were performed at 10 mK.

**Monolayer $MoS_2$ Device Campaign**

We first demonstrate the end-to-end autonomous experimentation framework using a monolayer $MoS_2$ device campaign. Overall, as shown in Fig. 3, the whole process aims to address scientific questions about which carrier density populates the upper band in $K$ and $K'$ valleys and the conduction-band SOC energy.

Constrained by the hypothesis-driven Agent Skills, the first measurement agent aims to answer a focused question: *what is the conducting gate-voltage window?* (Fig. 3a). It then decides to measure the $R_{xx}$ dependence of $V_{TG}$ and generates the measurement commands. After the measurement is completed, the analysis agent analyzes the gate dependence to determine the turn-on voltage by tracking the most negative value of $\frac{dR_{xx}}{dV_{TG}}$. Then the analysis agent evaluates the result against the acceptance criteria and decides whether to proceed to the next step, redo the gate-dependence measurement, or conclude with a low-quality device by skipping further measurements. With sufficient evidence in the previous step, the second measurement agent tries to answer a different question: *Where do quantum oscillations appear in the plane of $V_{TG}$ and $B$?* To address this question, the agent decides to perform the Landau fan measurement (Fig. 3b). In detail, it chooses the range of $V_{TG}$ starting from the turn-on voltage based on the previous-step outcome. After the measurement is finished, a new agent analyzes the data by performing the Fast Fourier Transform (FFT) against $1/B$ for each $V_{TG}$ and defines 12 provisional frequency branches and 2 provisional fundamental frequencies ($F_1$ and $F_2$) for further analysis (Supplementary Figure 1).

After these two rounds of agent loops, a new measurement agent (Fig. 3c) answers the question of *whether the FFT frequencies obtained at the previous step are reliable*. Specifically, the agent decides to sweep $B$ continuously at selected $V_{TG}$. Following this experiment, a new analysis agent analyzes the fundamental FFT frequency and compares it with the provisional fundamental frequency. After gathering this information, a new measurement agent focuses on addressing *what* *is the degeneracy of band corresponding to each fundamental FFT frequency*, which can be answered by performing the Hall resistance measurement shown in Fig. 3d. A new analysis agent follows this measurement and compares the Hall density with fundamental FFT frequency and concludes that there exists only one fundamental frequency ($F_1$) with degeneracy of 2 below $V_{TG} = 1.4\ V$, while the second fundamental frequency ($F_2$) starts from $V_{TG} = 1.4\ V$ with degeneracy of 2.

To conclude this autonomous agent workflow, a new agent provides the final assignment of the branches in the FFT map with four confidential levels: accepted, rejected, ambiguous, or unassigned. This agent concludes that only the lower conduction band is occupied below the carrier density $n \approx 3.4 \times 10^{12}\ cm^{-2}$, and the associated SOC splitting energy is 13 meV with the given effective mass [31] (Supplementary Figure 3).

To examine the robustness of this monolayer $MoS_2$ agent workflow, we repeat this autonomous agent campaign on a second, low-quality monolayer device. Very interestingly, after step 2 (Supplementary Figure 4), the agent fails to assign clear FFT peaks, but it still tries to scan $B$ at two $V_{TG}$ values. Because the results from this step are still negative, the agent skips the rest of the measurement and jumps to the last step to generate a report stating that no peak is accepted in the FFT map. This negative test case clearly demonstrates that our hypothesis-driven AI agent harness behaves more like a condensed-matter experimentalist than a simple deterministic experimental workflow.

**Bilayer $MoS_2$ Device Campaign**

Compared to monolayer $MoS_2$ devices, performing quantum transport measurements on bilayer $MoS_2$ devices poses a qualitatively harder challenge for AI agents. The phase diagram of bilayer $MoS_2$ devices is sensitive to both the carrier density ($n$) and electric field ($E$). For the bilayer $MoS_2$ measurements, as demonstrated in Fig. 4, the agents collectively address the scientific problem of *which layer and valley components are occupied in the plane of* $n$ *and* $E$.

As in the monolayer $MoS_2$ measurements, at each round of the experimental loop, the agent aims to answer a focused question through experimental measurements. For example, the first measurement agent concentrates on: *what is the ratio of top/bottom hBN thickness* (Fig. 4a). To resolve this question, the agent measures the $R_{xx}$ as a function of $V_{TG}$ and $V_{BG}$ at $B = 0$. Then, an analysis agent tracks the constant resistance in the contour plot and captures the ratio of top and bottom hBN thickness, followed by plotting the $R_{xx}$ as a function of $E$ and a provisional $n$.

*What are the thicknesses of top and bottom hBN?* To answer this second question, a new measurement agent subsequently measures the Hall resistance $R_{xy}$ at $B = \pm 6$ T and $V_{TG} = 2$ V by sweeping the $V_{BG}$ for calibration (Fig. 4b). After this measurement, an analysis agent fits the Hall density dependence with $V_{BG}$ and calculates the hBN thicknesses. For the third round, the agent now focuses on a new question: *where are layer-polarized and layer-shared boundaries in the plane of* $n$ *and* $E$. To resolve this issue, the agent measures the $R_{xx}$ map of $V_{TG}$ and $V_{BG}$ at $B = 6, 9, 13$ T. Then, a new analysis agent plots the data and extracts the phase boundaries by determining the point at which the strips from the quantum oscillation of $R_{xx}$ becomes diagonal in the $n - E$ phase space. It analyses the maps at different magnetic fields and concludes with a schematic showing the layer-polarized and layer-shared regions (Fig. 4c and Supplementary Figure 5). After this step, a final measurement agent answers: *what is the valley occupation in each region*? The Agent Skills have background knowledge that the degeneracy is 2 for the $K$ valley lower and upper bands and 12 for the $Q$ valley, while in the charge-shared region, the degeneracy doubles due to layer degeneracy. With this background knowledge, the agent decides to choose three electric fields that cut through three regions and measures the Landau fan diagram. After the measurements finish, a new agent extracts the fundamental frequency and assigns them to degeneracy and valleys (Fig. 4d). Finally, a new agent draws the phase diagram in the plane of $n$ and $E$ to show the layer and valley occupations (Fig. 4e). The autonomous agent campaign on a second bilayer device is included in Supplementary Figure 6.

**Discussion and Outlook**

In conclusion, we developed a multi-step agent harness for autonomously studying 2D materials devices. Using experimentally validated Agent Skills for monolayer and bilayer TMD devices, the agent can reliably run multi-step experiments end-to-end over many days within the FermiLink harness.

Three design choices appear essential for the reliability of experimental outcomes. First, the decision-tree structure of Agent Skills, which contains one narrow question per step with quantitative acceptance criteria, converts open-ended experimentation into a sequence of verifiable decisions. Second, multiple measurement and analysis agents working under the same file system allow a coherent experimental campaign to span many days via shared memory documents and experimental data. Third, the separation between the agent-writable surface and the hardware safety layer provides a hard safety gate for protecting both the valuable TMD devices and cryogenic equipment, increasing our confidence in using AI agents for quantum transport measurements.

Despite these insights, the limitations of this work are equally instructive. At this moment, for every new type of 2D device, human experts must collaborate with AI agents to develop robust Agent Skills for autonomous multi-step measurements using the existing dataset via dry runs (Fig. 1b). Thus, the autonomy demonstrated here primarily concerns the execution and adaptation of expert-defined experimental workflows, and it would be more appealing for AI agents to independently formulate measurement strategies from first principles. To this end, an important future direction is to develop AI agents that can construct and validate new Agent Skills from prior experimental datasets and a library of established workflows. As such a library expands across materials and device architectures, agents may be able to identify reusable measurement and analysis patterns while retaining safety constraints. Moreover, beyond performing human-defined experimental workflows, future agents could simultaneously propose scientific questions, design discriminating experiments, perform theoretical calculations for cross-validation, and search for large experimental parameter spaces for novel phases. While achieving broader scientific autonomy remains challenging, our work represents a step toward self-driving laboratories of quantum materials discovery [11,14,34].

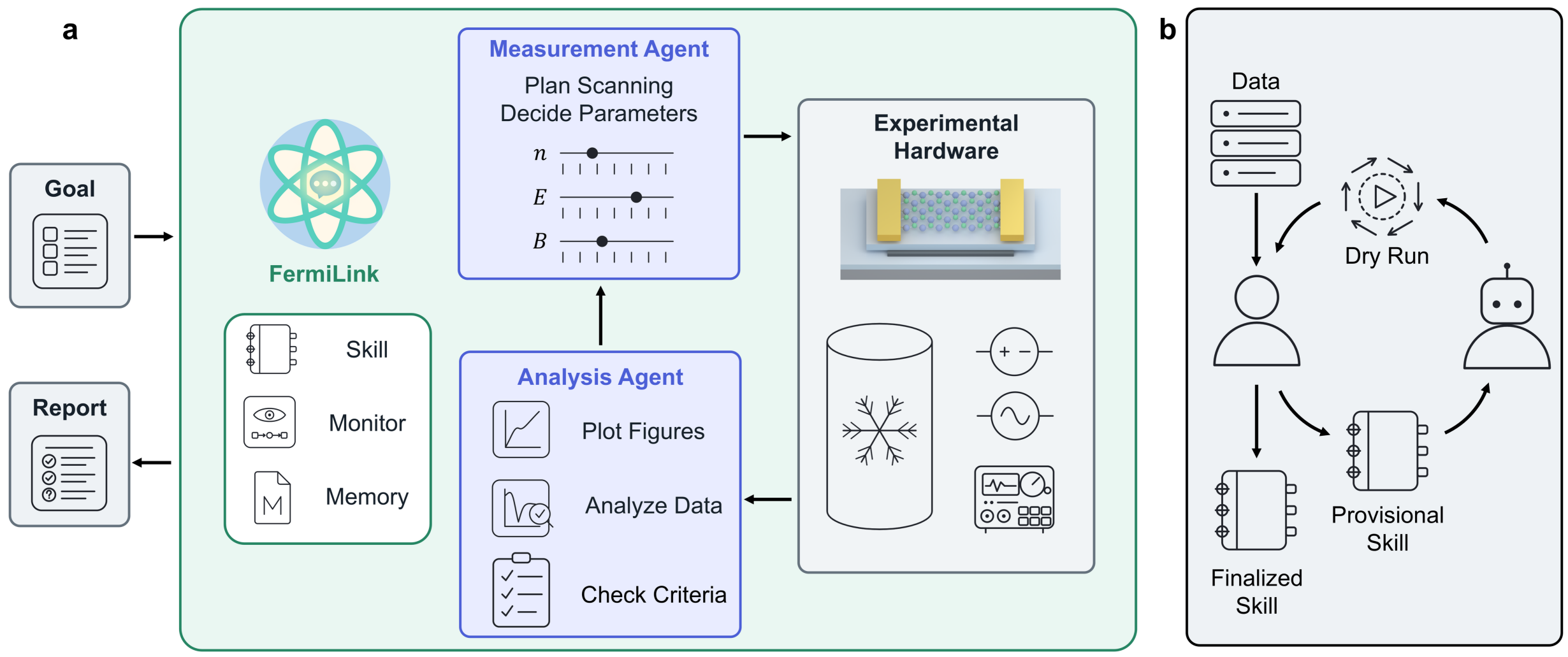

**Figure 1. Autonomous experimentation framework. a**, End-to-end agent operation. Given a user-supplied brief instruction, the FermiLink agent harness launches a measurement agent that plans the scan and decides the parameters (e.g., carrier density $n$, electric field $E$, and magnetic field $B$), followed by submitting the measurement to the experimental hardware through a parser-controlled acquisition interface. When the measurement finishes, an analysis agent post-processes the data and decides whether to proceed to the next step or skip to the final step based on the acceptance criteria. The measurement-analysis loop repeats until all sub-questions predefined in Agent Skills are answered and a final Report is generated. **b**, Development of hypothesis-driven Agent Skills. Starting from a measurement dataset and provisional Agent Skills, the AI agents *simulate* the experimental measurements and analysis against the dataset, while a human expert determines whether to further modify the Agent Skills for more robust performance. The converged Agent Skills are provided for the FermiLink harness for *real* experiments.

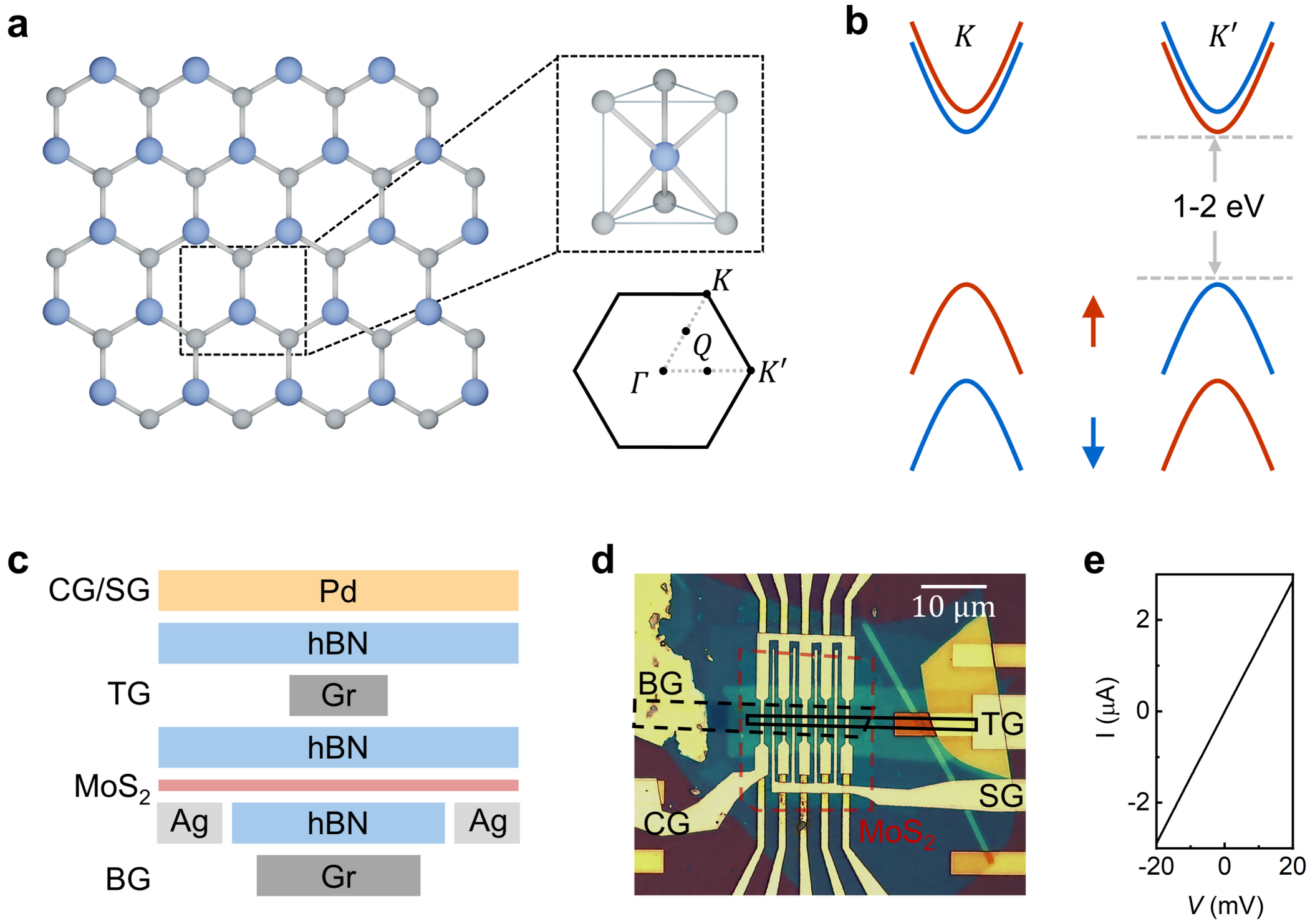

**Fig. 2 | $MoS_2$ devices and electronic structure. a**, Top view of the TMD lattice, with the transition metal (blue) and chalcogen (grey) sublattices. Right inset: the hexagonal Brillouin zone, indicating the $\Gamma$, $K$, and $K'$ points and the $Q$ valleys. **b**, Schematic band structure at the $K$ and $K'$ valleys of monolayer $MoS_2$. Red and blue denote spin-up and spin-down bands. **c**, Schematic cross-section of the dual-gated Hall bar device. The $MoS_2$ channel (red) is encapsulated in hBN, with graphite top (TG) and bottom (BG) gates, Pd contact/split gates (CG/SG), and Ag electrodes. **d**, Optical image of the device, with the $MoS_2$ flake outlined in red and the gate electrodes labeled as in **c**. Scale bar, 10 μm. **e**, Representative two-terminal $I$–$V$ characteristic measured at 10 mK.

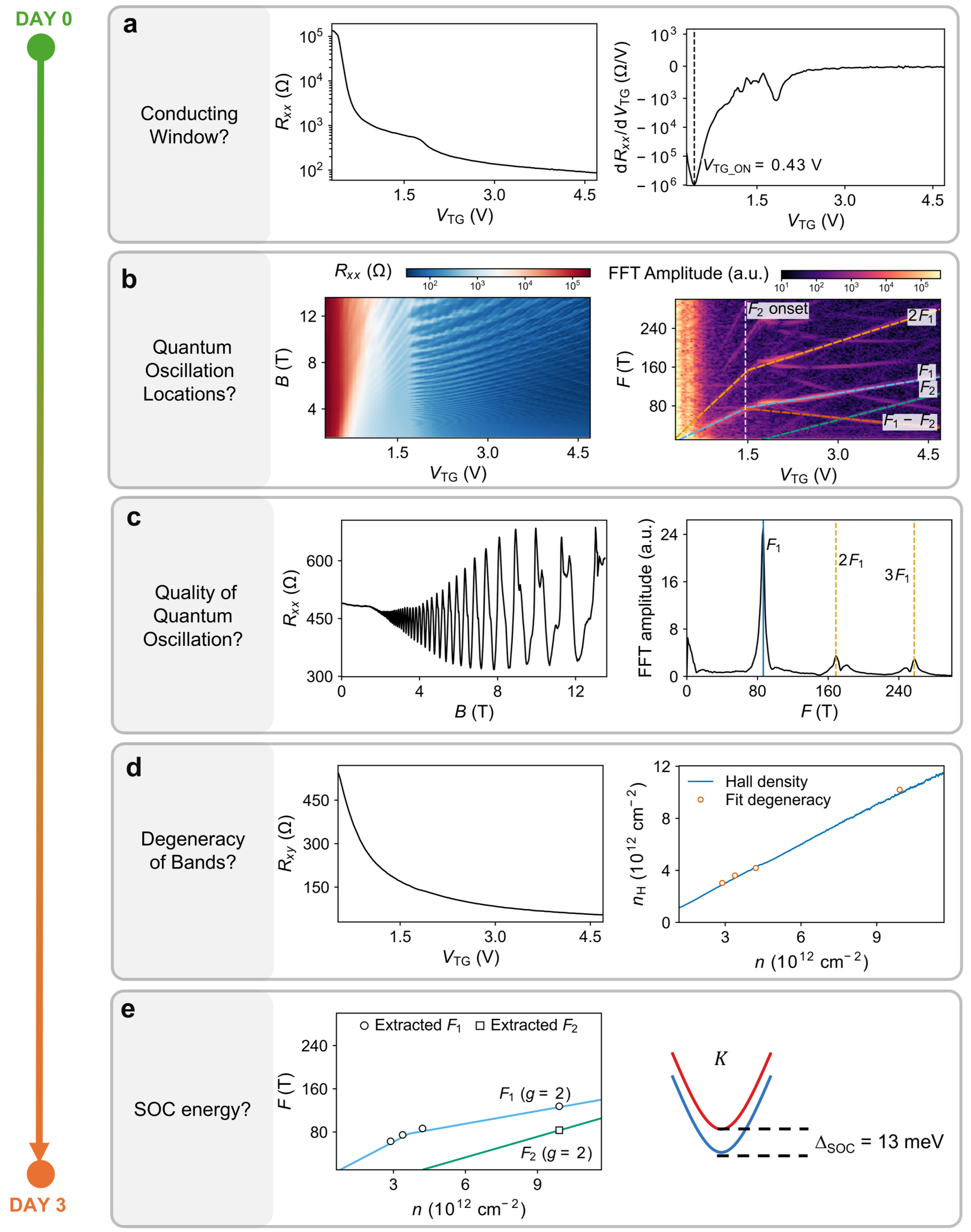

**Figure 3. Autonomous monolayer $MoS_2$ campaign.** Each row is one round of the experiment loop, labeled by the sub-question the agent set out to answer. The left figures present the raw data, and the right figures show the agent analysis. **a**, $R_{xx}$ as a function of $V_{TG}$ at $B = 0$. The turn-on voltage $V_{TG_ON} = 0.43$ V is marked by the black dotted line. **b**, $R_{xx}$ as a function of $V_{TG}$ and $B$ (left) and its gate-resolved FFT in $1/B$ (right). The provisional frequency branches ($F_1$, $F_2$, $2F_1$, $F_1 - F_2$) and the $F_2$ onset are marked. **c**, A representative continuous field sweep at $V_{TG} = 1.73\ V$ (left) and its FFT (right). The fundamental frequency $F_1$ and its harmonics $2F_1$ and $3F_1$ are marked. **d**, Hall resistance $R_{xy}$ at $B = 1$ T (left) and the Hall density $n_H$ as a function of $V_{TG}$ (right). The orange circles compare $n_H$ with the density calculated from the FFT frequencies after assigning the degeneracy of each band. **e**, Extracted frequencies $F_1$ and $F_2$ versus $V_{TG}$, each assigned degeneracy $g = 2$, yielding a conduction-band SOC splitting energy $\Delta_{SOC} \approx 13$ meV.

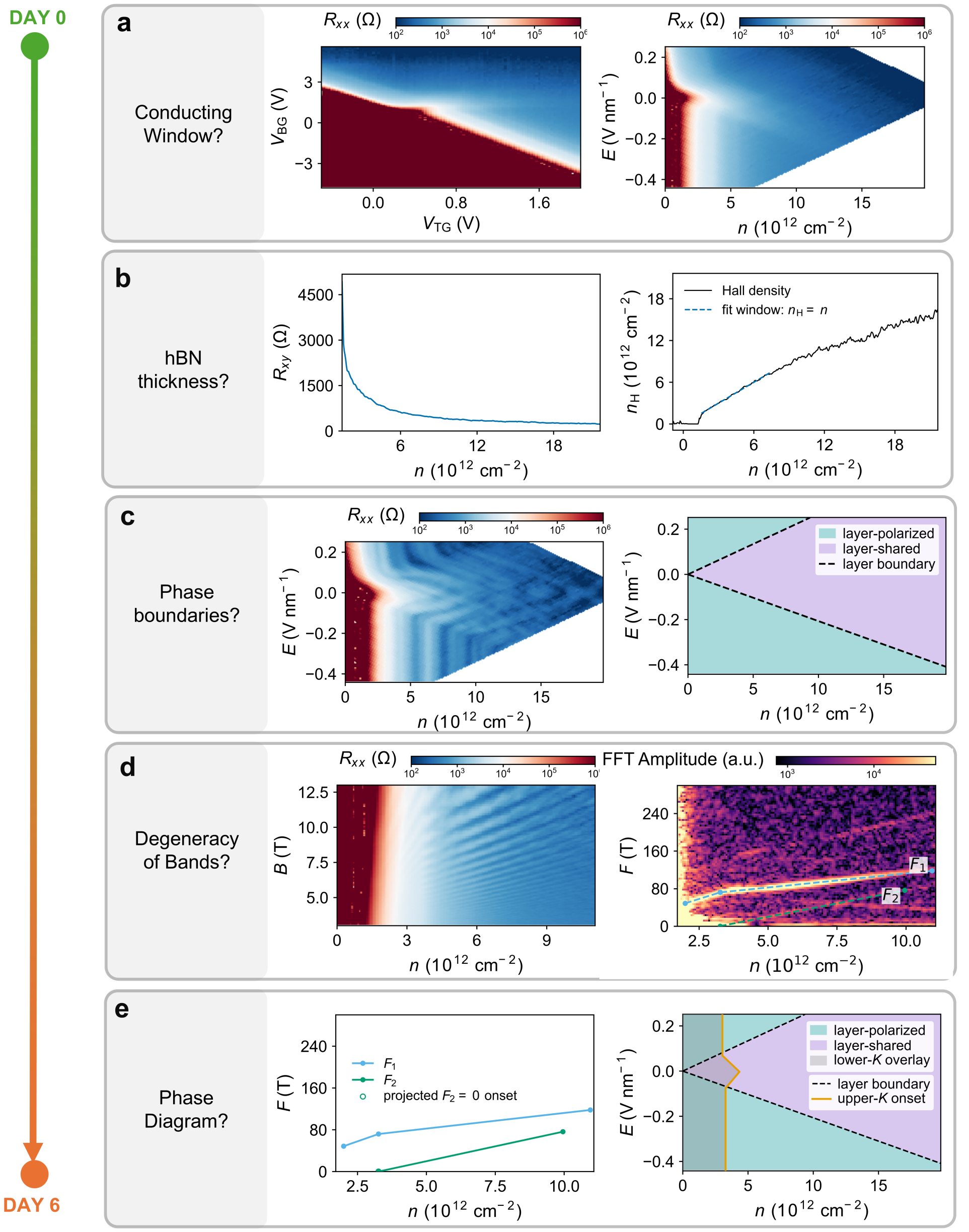

**Fig. 4. Autonomous bilayer MoS₂ campaign.** Each row is one round of the experiment loop, labeled by the sub-question the agent set out to answer. **a**, $R_{xx}$ as a function of $V_{TG}$ and $V_{BG}$ (left) and $n$ and $E$ (right). **b**, $R_{xy}$ at $B = 6$ T (left) and Hall density $n_H$ (right) versus $n$. The fit window used to calibrate the thickness of top and bottom hBNs is shown by the blue dashed line. **c**, Left: $R_{xx}$ contour map at $B = 13$ T. Right: provisional layer boundaries assigned by the agent. **d**, $R_{xx}$ as a function of $n$ and $B$ at $E = -0.275$ V/nm (left) and its normalized gate-resolved FFT (right). **e**, Top: fundamental frequencies $F_1$ and $F_2$ with degeneracy and valley assignments. Right: the final annotated phase diagram, showing the layer-polarized and layer-shared regions together with the valley boundaries.